\documentclass[a4paper,11pt]{article}
\pdfoutput=1 

\usepackage{jheppub} 

\usepackage[T1]{fontenc} 

\title{\boldmath Realization of electroweak baryogenesis by fourth generation fermions}

\author[a,1]{Hsiang-nan Li\note{Corresponding author.}}

\affiliation[a]{Institute of Physics, Academia Sinica,\\Taipei, Taiwan 115, Republic of China}

\emailAdd{hnli@phys.sinica.edu.tw}

\abstract{We demonstrate that the electroweak baryogenesis can be realized in the extended 
Standard Model with sequential fourth generation fermions (SM4). The solution to the 
coupled Dyson-Schwinger equations for the fermion and Higgs masses indicates that fourth 
generation quarks $t'$ and $b'$ with the Yukawa couplings above the threshold 
$g_Q^c\approx 9.1$ form condensates. This critical coupling, greater than the value 
$g_Y^f\approx 7$ at the ultraviolet fixed point of the renormalization-group evolution in 
the SM4, implies the existence of an electroweak symmetric phase at a high energy. The 
$t'$ and $b'$ Yukawa couplings evolve as the energy decreases, and exceed $g_Q^c$ at a 
lower scale. The Higgs potential with the dynamical symmetry breaking effect from heavy 
quark condensates plus the one-loop temperature-dependent correction from the heavy scalars 
formed by $t'$ and $b'$ quarks allow the first-order phase transition characterized by 
the ratio $\phi_c/T_c\approx 0.9$, where $\phi_c$ is the location of the Higgs potential 
minimum at the critical temperature $T_c$. Together with the baryon number violating 
sphaleron interaction inherent in the Standard Model and the enhanced $CP$ violation source 
from fourth generation quarks, the baryon asymmetry in the Universe can be achieved.}

\begin{document} 
\maketitle
\flushbottom

\section{Introduction}
\label{sec:intro}

We have performed dispersive analyses on the flavor structure of the Standard Model (SM)
in a series of publications recently \cite{Li:2023dqi,Li:2023yay,Li:2023ncg,Li:2024awx}. 
Sufficient clues for understanding the mass hierarchy and the distinct mixing patters of 
quarks and leptons have been accumulated, which suggest that the scalar sector could be 
stringently constrained by the internal consistency of SM dynamics. We were then motivated 
to explore the sequential fourth generation model as a natural and simple extension of the 
SM (hereafter abbreviated as SM4), to which no free parameters are added 
\cite{Li:2023fim,Li:2024xnl}. For example, the mass $m_{t'}\approx 200$ TeV 
($m_{b'}= 2.7$ TeV) of a fourth generation quark $t'$ ($b'$) was determined by the 
dispersion relation for neutral quark state mixing through box diagrams \cite{Li:2023fim}. 
It has been pointed out that $t'$ and $b'$ quarks with masses above a TeV scale form bound 
states in a Yukawa potential \cite{Hung:2009hy,Enkhbat:2011vp}. The contributions from 
$b'\bar b'$ scalars to the Higgs boson production via gluon fusion and to the Higgs decay 
into a photon pair were shown to be suppressed at least by a factor of $10^{-2}$ compared 
with the top quark ones \cite{Li:2023fim}. The impact on the oblique parameters from fourth 
generation quarks was also within experimental errors \cite{Li:2024xnl}. The above 
examinations elucidated why these superheavy quarks bypass current experimental constraints.

The origin of the baryon asymmetry in the Universe (BAU) is still not understood, and remains
as a challenge. Sakharov outlined three conditions required for the BAU \cite{Sakharov:1967dj}: 
baryon number violation, $C$ and $CP$ violation, and departure from thermal equilibrium.
Whether the BAU can be realized in the SM has been scrutinized in the 
literature \cite{Farrar:1993sp,Farrar:1993hn,Gavela:1993ts,Gavela:1994ds,Gavela:1994dt}. 
The SM contains the baryon number violating sphaleron interaction, provides the $CP$ violation 
source from the Kobayashi-Maskawa (KM) mechanism, and exhibits the electroweak phase transition 
(EWPT), which is, however, not strongly first-order \cite{Kajantie:1996mn,Rummukainen:1998as,
Csikor:1998eu,Aoki:1999fi}. The insufficient strength of the EWPT is magnificent enough for 
demanding the introduction of new physics. The baryogenesis in the SM4 has been 
extensively investigated, such as the dynamical electroweak symmetry breaking (EWSB) 
through heavy fermion condensates 
\cite{Holdom:1986rn,Bardeen:1989ds,Hill:1990ge,Elliott:1992xg,Mimura:2012vw}, 
the first-order EWPT \cite{Ham:2004xh,Carena:2004ha}
for attaining thermal non-equilibrium, and the enriched $CP$ violation sources 
\cite{Hou:2008xd}. Since the SM4 with fourth generation quarks heavier than a TeV scale is 
viable, and the previous works assumed the degenerate $t'$ and $b'$ masses slightly above 
the electroweak scale (about hundreds of GeV), it is worth exploring whether the SM4 with 
superheavy fourth generation quarks of different masses can produce the dynamical EWSB and 
satisfy Sakharov's three criteria. 

We will demonstrate that, contrary to the conclusions in the literature 
\cite{Fok:2008yg,Kikukawa:2009mu}, the electroweak baryogenesis can be achieved in the SM4.
Suppose that there exists a strongly interacting sector at a high energy, like the 
composite Higgs model described in \cite{Kaplan:1983fs}, whose detailed structure is not 
essential for our discussion. This type of models is first broken down to the 
$SU_L(2)\times U_Y(1)$ theory, i.e., the SM4 in the electroweak symmetric phase, at a
scale just below the compositeness scale. At such a high energy, the SM4 should have
reached the ultraviolet fixed point of the two-loop renormalization-group 
(RG) evolution observed in \cite{Hung:2009hy}, where the Yukawa couplings 
of all fourth generation fermions take the similar value $g_Y^f \approx 7$. We check 
whether the dynamical EWSB can be induced by fourth generation fermions at this fixed point 
by solving the coupled Dyson-Schwinger (DS) equations \cite{Dyson:1949ha,Schwinger:1951ex} 
for the fermion and Higgs masses \cite{Hung:2010xh}. It turns out that the critical coupling 
for heavy fermions to form condensates, which trigger the EWSB, is $g_Y^c \approx 8.0$, 
above the fixed-point value $g_Y^f \approx 7$. The relation $g_Y^c > g_Y^f$ implies that the 
electroweak symmetry is maintained just below the compositeness scale, and all fourth 
generation fermions stay massless. This is exactly the symmetric phase we postulated in our 
previous dispersive analyses on the neutrino mixing \cite{Li:2023ncg,Li:2024awx}: the 
disappearance of the neutrino mixing phenomenon in the symmetric phase serves as a high energy 
input to the associated dispersion relations, from which constraints on the neutrino masses 
and mixing angles were extracted. This symmetric phase also lays a solid ground for  
formulating the electroweak factorization of ultra-high energy scattering processes 
\cite{Chien:2018ohd}.

As the energy scale further goes down, the Yukawa couplings of fourth generation fermions 
begin to deviate away from the common fixed-point value following their individual 
RG evolution \cite{Hung:2009hy}. The $t'$ and $b'$ Yukawa 
couplings, with the masses $m_{t'}\approx 200$ TeV and $m_{b'}=2.7$ TeV at the electroweak 
scale \cite{Li:2023fim}, grow from the fixed-point value. In contrast, the Yukawa 
couplings of fourth generation charged lepton $L$ and neutrino $\nu_4$, with the masses 
$m_{L}=270$ GeV and $m_4=170$ GeV at the electroweak scale \cite{Li:2024xnl}, drop from 
the fixed-point value. The former then have a chance to exceed the critical coupling 
as the scale decreases, such that the heavy quark condensate 
$\langle\bar t't'+\bar b'b'\rangle<0$ can be established. We illustrate by solving the 
DS equations that the heavy quark condensate is formed as the heavy quark Yukawa coupling 
reaches $g_Q^c\approx 9.1$, which corresponds to a quark mass 1.6 TeV. Both $t'$ and $b'$ 
quarks can certainly arrive at this mass under the RG evolution in view of their 
aforementioned masses at the electroweak scale. The heavy quark condensate, 
giving rise to a quadratic term $\mu^2\phi^2/2$ with the mass parameter $\mu^2<0$ in 
the Higgs potential, breaks the electroweak symmetry, and all other particles get 
masses. Additional heavy scalars (or pseudoscalars) then appear as bound states of $t'$ 
and $b'$ quarks \cite{Bardeen:1989ds}, so the huge $t'$ Yukawa coupling is in fact never 
encountered during the RG evolution, and the unitarity is not an issue in our formalism. 
It will be affirmed that the quartic term in the Higgs potential can also be generated in 
the above effective theoretical approach \cite{Hung:2010xh,Frasca:2024ame}. Fourth 
generation leptons constitute to neither the condensate nor the dynamical EWSB, because 
the RG effect lowers their Yukawa couplings to the $O(1)$ level at the electroweak scale. 
This marks a distinction from the conclusion in \cite{Hung:2010xh}, where all fourth 
generation fermions form condensates.

Next we study the EWPT in the SM4, deriving the standard effective Higgs potential 
$V_{\rm eff}(\phi)$ \cite{Dolan:1973qd}, which is expressed as the sum of the tree-level 
potential $V_0(\phi)$ at zero temperature $T=0$, the one-loop Coleman-Weinberg correction 
$V_1(\phi)$ at zero temperature \cite{Coleman:1973jx} and the one-loop $T$-dependent 
correction $V_T(\phi)$. The trivial vacuum at $\phi=0$ and the nontrivial vacuum at 
$\phi=\phi_c$ for $V_{\rm eff}(\phi)$ are identified at various $T$, which become equal,
$V_{\rm eff}(0)=V_{\rm eff}(\phi_c)$, at the critical temperature $T=T_c$. The obtained 
ratio $\phi_c/T_c \approx 0.9$ roughly meets the criterion for the first-order phase 
transition $\phi_c/T_c \gtrsim 1$. The pivotal impact on $V_1(\phi)$ and $V_T(\phi)$
from the heavy scalars, i.e., the bound states of fourth generation quarks, will be 
highlighted; it deepens the nontrivial vacuum of the effective Higgs potential, such 
that a barrier between it and the trivial vacuum can be created by the $T$-dependent 
$V_T(\phi)$. The absence of the heavy scalars would render the EWPT second-order or 
crossover. The absence of fourth generation leptons would increase $T_c$ 
more than $\phi_c$, and decrease their ratio to $\phi_c/T_c \approx 0.8$. The above 
pinpoint how the involvement of fourth generation fermions makes the first-order EWPT, a 
scenario quite different from those proposed in the literature that usually rely on 
the extension of the scalar sector \cite{Chiang:2014hia,Fuyuto:2017ewj,Goncalves:2021egx,
Biekotter:2021ysx,Chatterjee:2022pxf,Carena:2022yvx,Zhang:2023jvh,Biekotter:2023eil,
Chaudhuri:2024vrd,Ramsey-Musolf:2024zex,Athron:2025iew}. The SM offers large baryon number 
violation \cite{tHooft:1976rip}, which is caused by the triangle anomaly through a 
nonperturbative effect from the vacuum structure of $SU(2)$ gauge theories. Fourth 
generation quarks can enhance $CP$ violation \cite{Hou:2008xd} according to the dimensional 
argument on the Jarlskog invariants \cite{Jarlskog:1985ht,Jarlskog:1985cw,Shaposhnikov:1986jp}. 
The strongly first-order EWPT in SM4 facilitates departure from thermal equilibrium. These 
features satisfy Sakharov's conditions, and manifest the mechanism responsible for the 
electroweak baryogenesis \cite{Cohen:1990py,Cohen:1990it,Kuzmin:1985mm,Cline:2006ts}.

The rest of the paper is organized as follows. We solve the DS equations for the
heavy fermion mass and the mass parameter $\mu^2$ in the Higgs potential in Sec.~\ref{dyn}. 
Two cases are undertaken, one with $\mu^2$ receiving the contributions from all 
fourth generation fermions, and another with $\mu^2$ receiving the contributions from 
fourth generation quarks $t'$ and $b'$. The former finds that the critical Yukawa 
coupling for establishing the heavy fermion condensates is higher than the one at the 
ultraviolet fixed point, and that an electroweak symmetric phase of the SM4 exists at a 
high energy. The latter certifies that $t'$ and $b'$ quarks with their masses above a TeV 
scale can form the condensate, which breaks the electroweak symmetry dynamically. The 
effective coupling between Higgs bosons and the heavy scalars is estimated by matching 
fourth generation quark and heavy scalar contributions to Higgs boson scattering in 
Sec.~\ref{fir}. This process occurs via a box diagram with internal fourth generation 
quarks at a high energy, and via a heavy scalar loop at the electroweak scale. The 
effective coupling can then be inferred by equating the two calculations at a high energy, 
and utilized to evaluate the one-loop potentials $V_1(\phi)$ and $V_T(\phi)$. 
The first-order EWPT in the SM4 is verified by deducing the ratio 
$\phi_c/T_c \approx 0.9$. We evince that the heavy scalars play important roles for 
intensifying the strength of the EWPT. Section~\ref{con} contains the conclusion.

\section{Dynamical symmetry breaking}
\label{dyn}

The Yukawa couplings at the ultraviolet fixed point of the two-loop RG evolution 
\cite{Hung:2009hy} do not depend on initial fermion masses at the electroweak scale, since 
their anomalous dimensions, being of ultraviolet origin, do not. All fourth generation 
fermions have the similar Yukawa coupling $g_Y^f\approx 7$ at the fixed point, for the 
gauge interactions, that differentiate fermion flavors, are negligible at a high energy 
scale; it is noticed that the anomalous dimensions are symmetric under the exchange of the 
Yukawa couplings for fourth generation quarks and leptons, once the gauge interactions are 
turned off \cite{Hung:2009hy}. We explore the mechanism of the EWSB by solving the DS 
equations \cite{Dyson:1949ha,Schwinger:1951ex}, a formalism 
developed in \cite{Barrios:1990xi,King:1990ym} and extended to higher precision in 
\cite{Frasca:2024ame} recently. As stated in the Introduction, we start with the SM4 in 
the symmetric phase \cite{Li:2023ncg,Li:2024awx}, where the fixed point of 
the RG evolution is claimed to be located. It will be shown 
that the critical Yukawa coupling $g_Y^c\approx 8.0$ for forming fermion condensates is 
larger than the fixed-point value $g_Y^f\approx 7$, so the symmetric phase does 
exist, as the energy scale descends from the compositeness scale.

\begin{figure}[tbp]
\centering 
\includegraphics[width=.65\textwidth]{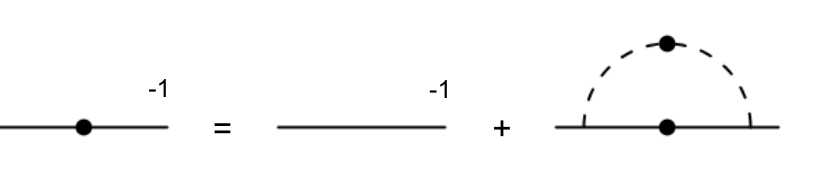}

\caption{\label{fig0}
DS equation for a quark propagator under the 
quenched and rainbow approximations, where the dots denote the
dynamically generated masses.}
\end{figure}

We thus take a common Yukawa coupling $g_Y$ for all four fourth generation fermions,
and write the DS equation for a quark propagator under the 
quenched and rainbow approximations \cite{Higashijima:1983gx,Roberts:1994dr} as
\begin{eqnarray}
-i[\not p-m(p^2)]=-i(\not p-m_0)-2(-ig_Y)^2\int\frac{d^4q}{(2\pi)^4}
\frac{i}{(p-q)^2-\mu^2(q^2)}
\frac{i[\not q+m(q)]}{q^2-m^2(q^2)},\label{1}
\end{eqnarray}
with the dynamically generated quark mass $m(q^2)$ (scalar mass squared $\mu^2(q^2)$). 
The above equation is graphically described in Fig.~\ref{fig0}. The bare 
quark mass $m_0$, appearing in the leading-order propagator, will be set to zero to 
accentuate the dynamical mass generation. A quark can emit a $\phi^0$ or $\phi^\pm$ 
boson (e.g., $t'\to t'\phi^0$ or $t'\to b'\phi^+$), which contributes equally to the 
second term on right-hand side of Eq.~(\ref{1}). This explains the presence of the 
coefficient 2. In principle, there should be an upper bound of the integration variable 
$q^2$, which is below the compositeness scale. Otherwise, contributions from new degrees 
of freedom need to be included into Eq.~(\ref{1}). We will elaborate that the extra 
term $\mu^2(q^2)$ compared to the corresponding equation in \cite{Hung:2010xh} grants 
the automatic emergence of the upper bound for $q^2$, as the EWSB is triggered. Therefore, 
an artificial upper bound like the one introduced in \cite{Hung:2010xh} is not necessary in 
our formulas. QCD corrections can be taken into account, but are not expected 
to modify our conclusions dramatically \cite{Barrios:1990xi}. For comparison, 
we mention Ref.~\cite{Mavromatos:2022heh}, which addressed the dynamical fermion mass 
generation by means of interactions with pseudoscalars.

We perform the trace of Eq.~(\ref{1}) to isolate the part for the quark 
mass, and then convert it into the Euclidean space through the Wick rotation, together 
with the transformation $p^0\to ip^0$. The resultant expression is given by
\begin{eqnarray}
m(p^2)=-2g_Y^2\int\frac{d^4q}{(2\pi)^4}\frac{1}{(p-q)^2+\mu^2(q^2)}
\frac{m(q^2)}{q^2+m^2(q^2)},\label{sdm}
\end{eqnarray}
where the dynamically generated $\mu^2(q^2)$ is related to 
the fermion condensate, i.e., the tadpole contribution, 
\begin{eqnarray}
\mu^2(p^2)=-(2N_C+2)\frac{2g_Y^2m(p^2)}{p^2+m^2(p^2)}
\int\frac{d^4q}{(2\pi)^4}\frac{m(q^2)}{q^2+m^2(q^2)},\label{sdu}
\end{eqnarray}
with the number of colors $N_c=3$. The coefficient $2N_c+2$ collects the contributions
from two heavy quarks $t'$ and $b'$ and two heavy leptons $L$ and $\nu_4$, which all 
have the common Yukawa coupling $g_Y$. The above equation can be understood in the viewpoint 
of an effective field theory, and has been also formulated in the Euclidean space. It 
approaches the corresponding equation in \cite{Hung:2010xh} in the zero-momentum limit 
$p\to 0$.

We write Eqs.~(\ref{sdm}) and (\ref{sdu}) as
\begin{eqnarray}
m(p^2)&=&-\frac{g_Y^2}{4\pi^3}\int_0^\infty q^2dq^2\int_0^\pi
\frac{\sin^2\psi d\psi}{q^2-2qp\cos\psi+p^2+\mu^2(q^2)}
\frac{m(q^2)}{q^2+m^2(q^2)},\label{dmp}\\
\mu^2(p^2)&=&-\frac{g_Y^2}{\pi^2}\frac{m(p^2)}{p^2+m^2(p^2)}
\int_0^\infty q^2dq^2\frac{m(q^2)}{q^2+m^2(q^2)}.\label{dup}
\end{eqnarray}
The integral over the polar angle $\psi$ gives
\begin{eqnarray}
f(p^2,q^2)\equiv\frac{1}{\pi}\int_0^\pi\frac{\sin^2\psi d\psi}{A-\cos\psi}
\approx\left\{\begin{array}{ccc}
   A-\sqrt{A^2-1},& A>1\\
   A,& -1<A<1,\\
   A+\sqrt{A^2-1},& A<-1,\\
   \end{array} \right.\label{ang}
\end{eqnarray}
with the function $A=[p^2+q^2+\mu^2(q^2)]/(2pq)$. Equation~(\ref{dmp}) reduces to
\begin{eqnarray}
m(p^2)&=&-\frac{g_Y^2}{4\pi^2}\int_0^{\infty} q^2dq^2\frac{f(p^2,q^2)}{2pq}
\frac{m(q^2)}{q^2+m^2(q^2)}.
\label{dmp2}
\end{eqnarray} 

We first draw some basic features of the unknowns $m(p^2)$ and $\mu^2(p^2)$ from the 
coupled integral equations~(\ref{dup}) and (\ref{dmp2}). A physical solution for $m(p^2)$ 
respects the positivity $m(p^2)\ge 0$, which implies $\mu^2(p^2)\le 0$ according to 
Eq.~(\ref{dup}). For $\mu^2(p^2)= 0$, we have $A\ge 0$ and $f(p^2,q^2)\ge 0$ based on 
Eq.~(\ref{ang}). The rest of the integrand $m(q^2)/[q^2+m^2(q^2)]$ in Eq.~(\ref{dmp2}) 
is positive. Hence, the right-hand side of Eq.~(\ref{dmp2}) is negative with 
the overall minus sign. However, the left-hand side of Eq.~(\ref{dmp2}) is positive, leading 
to contradiction. The only possibility is $m(p^2)=0$, which yields $\mu^2(p^2)=0$ from 
Eq.~(\ref{dup}); namely, the $SU_L(2)\times U_Y(1)$ electroweak symmetry is not broken. As 
long as $|\mu^2(q^2)|$ is tiny owing to a diminishing $g_Y$, the right-hand side of 
Eq.~(\ref{dmp2}), being negative, cannot be equal to the left-hand side. The coupling $g_Y$ 
must be finite, or equivalently, $\mu^2(p^2)<0$ must be sizable to overcome
$p^2+q^2$ in $A$, so that the right-hand side of Eq.~(\ref{dmp2}) can develop a positive 
result to match the left-hand side $m(p^2)>0$. This defines the critical coupling $g_Y^c$ 
that we will search for. Equation~(\ref{dmp2}) also indicates that $m(q^2)$ must drop to 
zero at a finite $q^2$; if $m(q^2)$ extends to infinite $q^2$, we can pick up $m(p^2)$ at a 
sufficiently large $p^2>-\mu^2(q^2)$ (a physical $-\mu^2(q^2)$ is finite), such that $A$ 
is positive, $f(p^2,q^2)$ is positive, and the right-hand side of Eq.~(\ref{dmp2}) is 
negative. Again, the left-hand side $m(p^2)$ is positive, ending up with contradiction.

\subsection{The symmetric phase}

The variable changes $p^2=m^2(0)x$ and $q^2=m^2(0)y$ recast Eq.~(\ref{dmp2}) into
\begin{eqnarray}
M(x)&=&-\frac{g_Y^2}{8\pi^2}\int_0^{\infty} ydy\frac{f(x,y)}{\sqrt{xy}}
\frac{M(y)}{y+M^2(y)},
\label{dmp3}
\end{eqnarray}
with the scaled quark mass $M(y)=m(y)/m(0)$, and define the dimensionless constant
$\tilde{\mu}^2$ via Eq.~(\ref{dup}),
\begin{eqnarray}
\tilde{\mu}^2\equiv\frac{\mu^2(0)}{m^2(0)}=-\frac{g_Y^2}{\pi^2}
\int_0^\infty dy\frac{M(y)y}{y+M^2(y)},\label{du3}
\end{eqnarray}
that appears in $f(x,y)$. It is easy to validate $M'(0)<0$ using Eqs.~(\ref{ang}) and 
(\ref{dmp3}), i.e., that $M(x)$ descends with $x$ at least in the low $x$ region. 
Together with the previous argument on the vanishing of $M(x)$ at finite $x$, we propose 
the ansatz for the unknown
\begin{eqnarray}
M(x)=\exp(-a x)\theta(b-x),\label{ans}
\end{eqnarray}
where the parameters $a$ and $b$ will be fixed below. It is basically a step 
function with the bound $x\le b$, modulated by a monotonically decreasing exponential 
with $a> 0$. Because Eq.~(\ref{ans}) is a simple ansatz, our result for the critical 
Yukawa coupling $g_Y^c$ should be regarded as a conservative one. It suffices for our 
purpose of demonstrating the existence of $g_Y^c$.

\begin{figure}[tbp]
\centering 
\includegraphics[width=.45\textwidth]{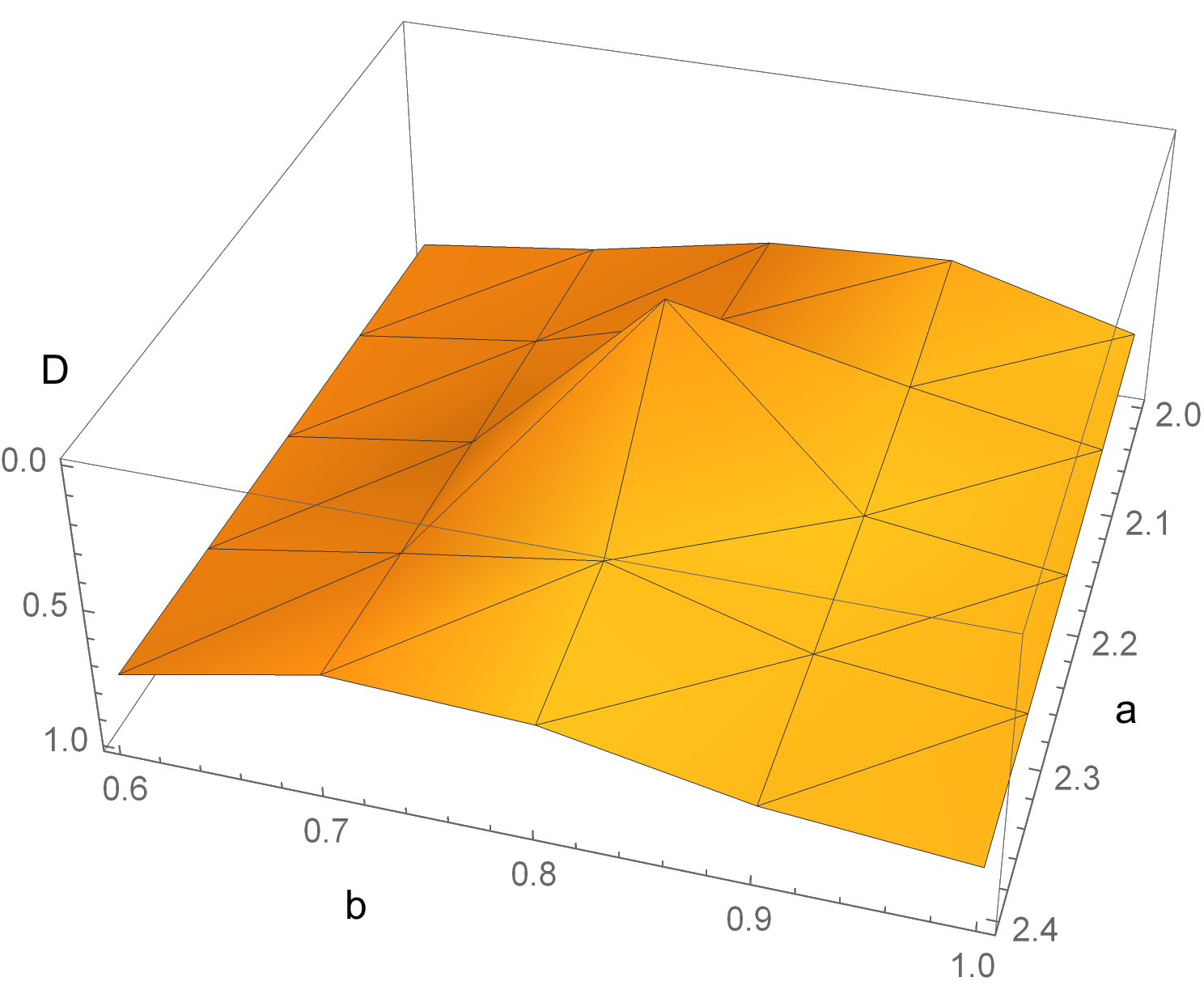}
\hfill
\includegraphics[width=.45\textwidth]{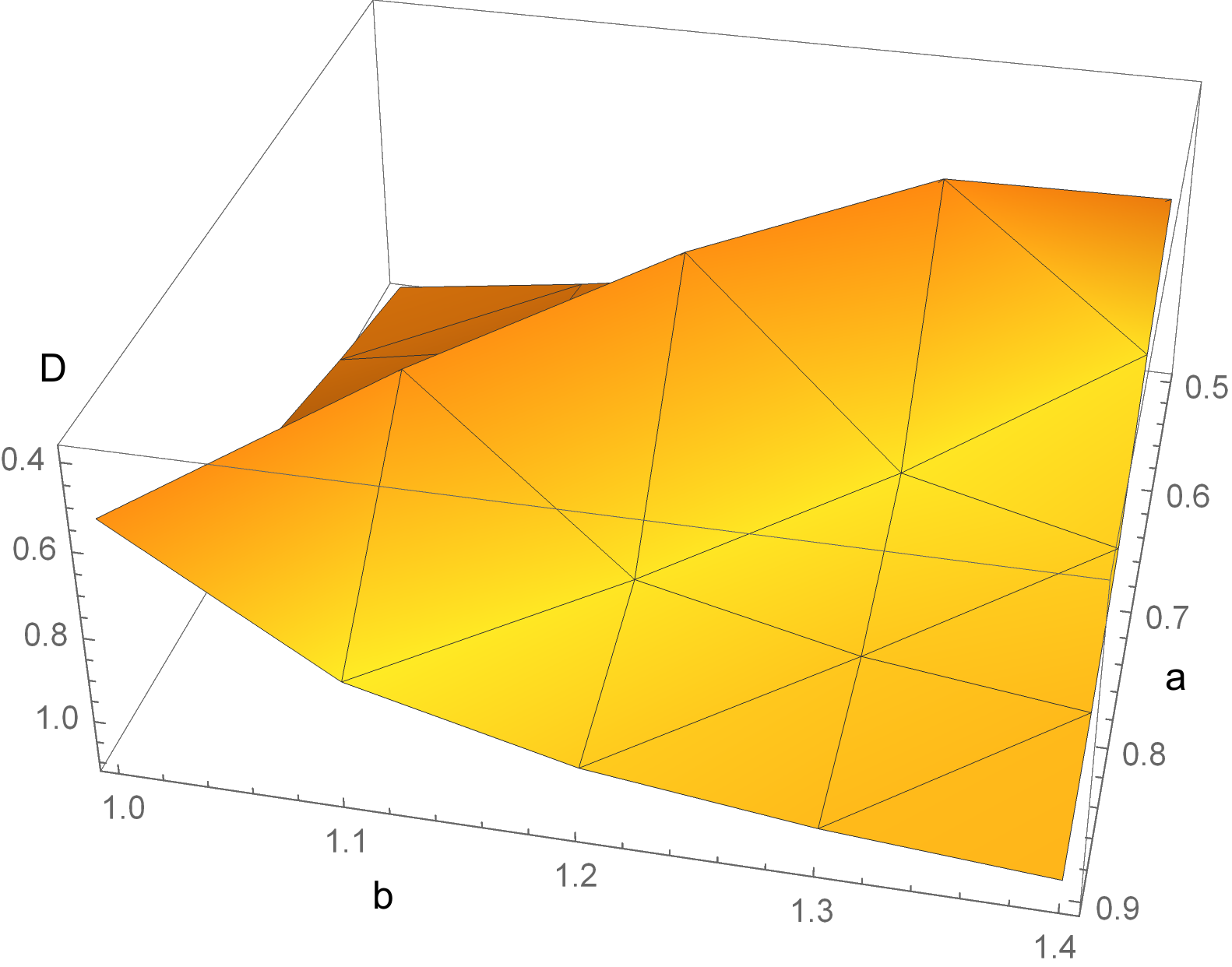}

(a)\hspace{8.0cm}(b)

\caption{\label{fig1}
(a) Dependence of the deviation $D$ in Eq.~(\ref{dev}) on $a$ and $b$
for $g_Y=15$, where the minimum located at $a=2.2$ and $b=0.8$ is displayed as 
a peak for clarity. (b) Same as (a) but for $g_Y=10$ with the location of the minimum
at $a=0.70$ and $b=1.2$.}
\end{figure}

We employ the two constraints from Eq.~(\ref{dmp3}) at the endpoints
$x\approx 0$ and $x\approx b$, which are chosen as $x=\epsilon=10^{-5}$ and
$x=0.9b$ specifically. It has been confirmed that other choices of $x$ do not alter
our solutions. The two parameters $a$ and $b$ are then acquired by minimizing the 
deviation
\begin{eqnarray}
D&=&\left|\frac{g_Y^2}{8\pi^2}\int_0^{b} ydy\frac{f(\epsilon,y)}{\sqrt{\epsilon y}}
\frac{M(y)}{y+M^2(y)}+M(\epsilon)\right|\nonumber\\
& &+\left|\frac{g_Y^2}{8\pi^2}\int_0^{b} ydy\frac{f(0.9b,y)}{\sqrt{0.9b y}}
\frac{M(y)}{y+M^2(y)}+M(0.9b)\right|,\label{dev}
\end{eqnarray}
with the ansatz in Eq.~(\ref{ans}) being inserted. If the fitted $a$ and $b$
render the right-hand side of Eq.~(\ref{dmp3}) reproduce the normalization $M(\epsilon)=1$, 
we will accept the solution corresponding to the minimum of Eq.~(\ref{dev}). This criterion 
is very loose apparently. It is the reason why our result represents a conservative one, 
and the actual critical Yukawa coupling is expected to be higher than the one obtained here. 
The deviation $D$ for the coupling $g_Y=15$ ($g_Y=10$) is presented in Fig.~\ref{fig1} 
as a typical example, where the minimum located at $a=2.2$ and $b=0.8$ 
($a=0.70$ and $b=1.2$), is displayed as a peak for clarity. It is trivial to 
verify $M(\epsilon)\approx 1.0$ in both cases, in line with our criterion. In other words, the 
minimal deviation in Fig.~\ref{fig1} is mainly attributed to the distinction between the 
two sides of Eq.~(\ref{dmp3}) near the high end of $x$.

By lowering the $g_Y$ value gradually, we check where a nontrivial solution ceases to 
exist. The minima of the deviation $D$ are less sharp, and
shift toward smaller $a$ with the decrease of $g_Y$, a tendency which has been
revealed in Fig.~\ref{fig1}. At the same time, the fit quality deteriorates; the height 
of the peak in Fig.~\ref{fig1}(a) for $g_Y=15$ is close to $D\approx 0$, compared with 
$D\approx 0.4$ in Fig.~\ref{fig1}(b) for $g_Y=10$. Eventually, a minimum of $D$, i.e., 
a solution to Eqs.~(\ref{dmp3}) and (\ref{du3}), disappears from the region $a> 0$. 
It turns out that solutions can be identified and the condition $M(\epsilon)\approx 1$ 
holds down to $g_Y=8.0$, with the fitted parameters $a=0.23$ and $b=0.91$. As 
$g_Y=7.9$, the minimum occurs at $a=0.04$, just adjacent to the boundary $a=0$, 
and $b=1.2$. When the coupling comes to $g_Y=7.8$, no minimum is found, and the 
inputs $a=0$ and $b=1.5$ (corresponding to the minimum in the variation of 
$b$) gives $M(\epsilon)=0.82$, in violation of our conservative criterion.  
We thus infer the critical Yukawa coupling $g_Y^c\approx 8.0$ around the ultraviolet fixed 
point of the SM4, greater than the result $\sqrt{2}\pi=4.4$ derived in \cite{Hung:2010xh}. 
The above examination hints that the simple ansatz in Eq.~(\ref{ans}) works for 
determining the critical Yukawa coupling from the DS equations. We compute the scaled mass 
parameters ${\tilde\mu}^2=-0.46$ and ${\tilde\mu}^2=-2.0$ from the outputs of $M(x)$ for 
the Yukawa couplings $g_Y=8.0$ and $g_Y=10$, respectively. These outcomes exhibit the 
desired correlation between $g_Y$ and $-{\tilde \mu}^2$, i.e., the strength of the 
condensate, which is supposed to grow with $g_Y$.

Our investigation declares that heavy fermions must have masses at least above a TeV 
scale ($g_Y=8.0$ is equivalent to a fermion mass $g_Yv/\sqrt{2}\approx 1.4$ TeV for the 
vacuum expectation value (VEV) $v=246$ GeV) in order to form condensates. The critical 
coupling $g_Y^c\approx 8.0$ is larger than the fixed-point value $g_Y^f\approx 7$ 
\cite{Hung:2009hy}, suggesting that fourth generation fermions cannot form condensates 
and the electroweak symmetry is maintained at a high energy right below the compositeness 
scale. This observation supports our assumption on the presence of the electroweak symmetric 
phase for the SM \cite{Li:2023ncg,Li:2024awx}, which is essential for arguing about the 
disappearance of the fermion mixing phenomena based on the unitarity of the 
Cabibbo-Kobayashi-Maskawa and the Pontecorvo–Maki–Nakagawa–Sakata matrices. The 
diminishing mixing amplitudes in the symmetric phase were then taken as the inputs to 
the relevant dispersion relations, from which the dispersive constraints on the fermion 
masses and mixing angles were constructed \cite{Li:2023ncg,Li:2024awx}. At last, 
the parameter ${\tilde\mu}^2=-0.46$ under the relations $m_H^2=-2\mu^2(0)$ and 
$\mu^2(0)=\tilde{\mu}^2m^2(0)$ corresponds to the Higgs mass $m_H\approx 1.3$ TeV, which is  
not very different from $m_H^*=1.44$ TeV at the ultraviolet fixed point 
\cite{Hung:2009hy}.

\subsection{Heavy quark condensates}

\begin{figure}[tbp]
\centering 
\includegraphics[width=.55\textwidth]{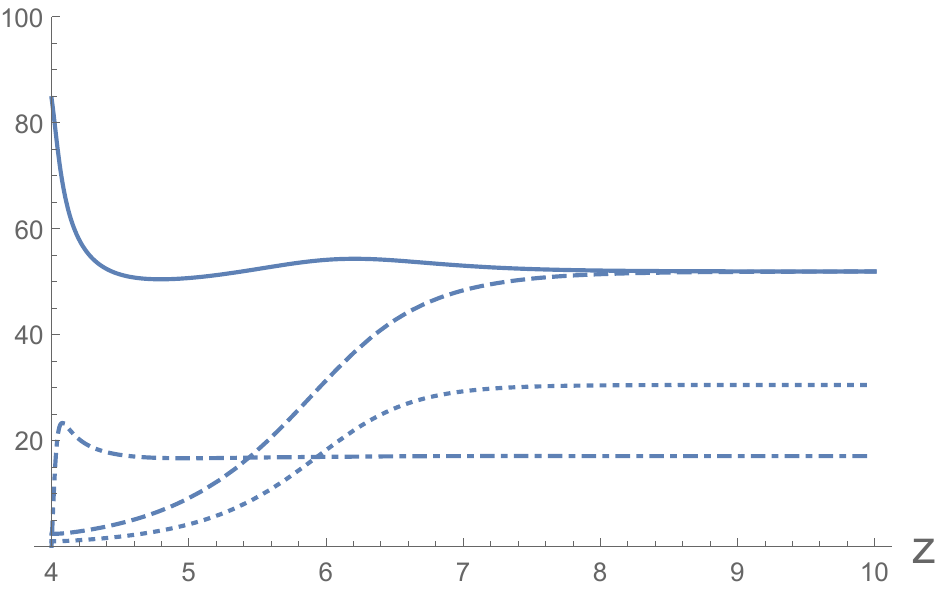}

\caption{\label{figY}
The RG evolutions of the squared Yukawa couplings $g_Q^2$ (solid line), $g_L^2$ (dashed
line) and $g_t^2$ (dotted line), and of the Higgs quartic coupling $\lambda$ (dot-dashed line)
with $z\equiv\ln(\mu/m_Z)$.}
\end{figure}

As the energy is lowered more, the Yukawa couplings of fourth generation fermions begin to 
run away from the ultraviolet fixed point following the two-loop RG evolution of the SM4 
\cite{Hung:2009hy}. With the $t'$ quark ($b'$ quark, charged lepton $L$, the neutrino 
$\nu_4$) mass about 200 TeV (2.7 TeV, 270 GeV, 170 GeV) at the electroweak scale, the 
Yukawa couplings of fourth generation quarks would increase from $g_Y^f\approx 7$, while 
those of fourth generation leptons decrease. As the energy drops to the so-called 
electroweak symmetry restoration scale, the former become large enough for establishing 
heavy quark condensates. It will be shown that the critical coupling 
$g_Q^c\approx 9.1$ in this case is also a bit higher than $g_Y^f$, so $t'$ and $b'$ 
quarks still have the similar Yukawa couplings at the symmetry restoration scale. 

To visualize the aforementioned evolutions, we solve the RG 
equations in Ref.~\cite{Hung:2009hy} for the squared Yukawa couplings $g_Q^2$, $g_L^2$ 
and $g_t^2$ of fourth generation quarks, fourth generation leptons and top quarks, 
respectively, and for the Higgs quartic coupling $\lambda$. We switch off the three gauge 
couplings in the anomalous dimensions in Eqs.~(2)-(8) of \cite{Hung:2009hy} for simplicity. 
The running in the dimensionless variable $z\equiv\ln(\mu/m_Z)$ as defined in 
\cite{Hung:2009hy}, $\mu$ being the renormalization scale and $m_Z$ being the $Z$ boson 
mass, is illustrated in Fig.~\ref{figY}. All the fixed-point results $g_Q^2=g_L^2\approx 52$, 
$g_t^2\approx 31$ and $\lambda\approx 17$ agree with those in \cite{Hung:2009hy}, 
confirming that the effects of the gauge couplings are indeed minor.

Figure~\ref{figY} discloses clearly that the Yukawa coupling $g_Q$ of
fourth generation quarks ascends with the decrease of $z$, despite being up-and-down 
slightly, from the fixed point around $z\approx 8$. We terminate the downward 
evolution at $z\approx 4$, where $g_Q$ reaches the critical coupling $g_Q^c\approx 9.1$, 
and the heavy-quark condensates are formed. On the contrary, $g_L$ descends with the 
decrease of $z$ from the common fixed-point value, and approaches the $O(1)$ coupling
at the electroweak scale. The behavior of $g_t$ for top quarks is similar 
to that of $g_L$. We remind that there is freedom to define the variable $z$; $m_Z$ in the 
argument of $\ln(\mu/m_Z)$ can be replaced by other electroweak scales, 
because a RG equation is formulated via the derivative 
$\mu d/d\mu=d/d\ln(\mu/{\rm EW\; scale})$. Therefore, the scales which characterize the 
features of the evolutions, such as the scale for the condensate formation and the location
of the fixed point, should be understood in terms of orders of magnitude.

Based on the above elucidation, we handle the DS equations (\ref{sdm}) 
and (\ref{sdu}) for fourth generation quarks in the same manner, but with the coefficient 
$2N_c+2$ in the latter being replaced by $2N_c$. Equation~(\ref{du3}) then turns into  
\begin{eqnarray}
\tilde{\mu}^2=-\frac{3g_Y^2}{4\pi^2}\int_0^\infty dy\frac{M(y)y}{y+M^2(y)}.
\label{sdu1}
\end{eqnarray}
The critical Yukawa coupling $g_Q^c$ at the symmetry restoration scale is expected to be 
greater than $g_Y^c$ in order to compensate the reduction of the coefficient from $2N_c+2$ 
to $2N_c$, but the distinction ought to be moderate. 
We solve for the unknown $M(x)$ at $g_Y=9.1$ from Eqs.~(\ref{dmp3}) and (\ref{sdu1}) by 
minimizing the deviation $D$ in Eqs.~(\ref{dev}), and get the fitted parameters 
$a=1.1$ and $b=0.14$, and the scaled mass parameter $\tilde\mu^2=-0.69$. As 
$g_Y=9.0$, the minimum of the deviation appears at $a=0.03$, just near the boundary 
$a=0$, and $b=1.4$. As the coupling goes down to $g_Y=8.9$, no minimum is 
identified, and the inputs $a=0$ and $b=1.5$ (corresponding to the minimum in the 
variation of $b$) yield $M(\epsilon)=0.88$, which does not meet our conservative 
criterion. We conclude the critical Yukawa coupling $g_Q^c\approx 9.1$ around the symmetry 
restoration scale in the SM4. 

The Yukawa couplings of a $t'$ quark with the mass $m_{t'}\approx 200$ TeV and of a $b'$ 
quark with the mass $m_{b'}= 2.7$ TeV at the electroweak scale can reach $g_Q^c$ definitely 
\cite{Li:2023fim} as manifested 
in Fig.~\ref{figY}. Fourth generation leptons with the masses $m_L\approx 270$ 
GeV and $m_4\approx 170$ GeV \cite{Li:2024xnl} at the electroweak scale do not contribute a 
condensate. We advocate that the mass parameter $\mu^2$ receives the contribution solely 
from fourth generation quarks $t'$ and $b'$ in our scenario for the EWSB. The value 
$-\tilde\mu^2\approx 0.69$ tells that the Higgs mass and the fourth generation quark 
masses are both of $O(1)$ TeV around the symmetry restoration scale.

\begin{figure}[tbp]
\centering 
\includegraphics[width=.35\textwidth]{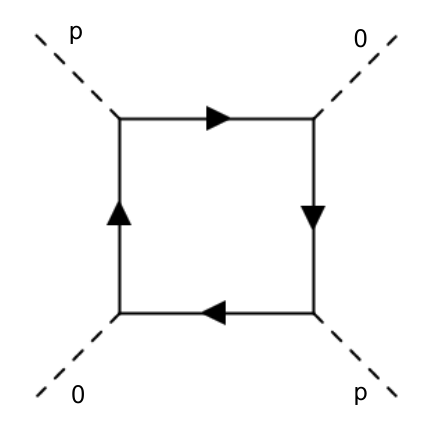}
\hfill
\includegraphics[width=.3\textwidth]{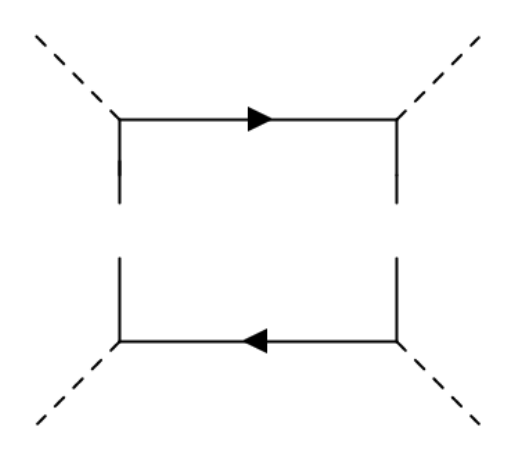}
\hfill
\includegraphics[width=.3\textwidth]{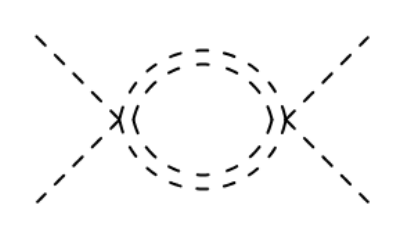}

(a)\hspace{4.7cm}(b)\hspace{4.7cm}(c)

\caption{\label{fig2}
(a) Box diagram with internal heavy quark lines. (b) Insertion of the Fierz transformation.
(c) Effective description for (a) at the electroweak scale, where the double dashed line
represents a heavy scalar $\eta$.}
\end{figure}

The quartic term of the Higgs potential can be generated by the heavy quark condensates 
too \cite{Hung:2010xh,Frasca:2024ame}. Consider the box diagram with 
four external scalars and internal heavy quark lines in Fig.~\ref{fig2}(a). Inserting the 
Fierz transformation for the fermion and color flows as depicted in Fig.~\ref{fig2}(b), we 
extract the four-quark condensate for the effective quartic scalar interaction, 
i.e., the self-coupling of Higgs bosons 
\begin{eqnarray}
-i\lambda&=&-2\left(\frac{g_Q^c}{\sqrt{2}}\right)^4\frac{4N_c}{m^2(0)}\int\frac{d^4q}{(2\pi)^4}
\left[\frac{m(q^2)}{q^2-m^2(q^2)}\right]^2,\label{la0}
\end{eqnarray}
where the minus sign on the right-hand side is associated with the fermion 
loop, the coefficient 2 counts the contributions from the internal $t'$ and $b'$ quarks,
and the factor $4N_c$ comes from the traces of the fermion and color flows. 
The substitution $q^0\to iq^0$ and the insertion of the output $M(y)$ for the 
Yukawa coupling $g_Q^c=9.1$ give
\begin{eqnarray}
\lambda&=&\frac{N_cg_Q^{c4}}{8\pi^2}\int_0^\infty ydy\left[\frac{M(y)}{y+M^2(y)}\right]^2
\approx 19.\label{lam}
\end{eqnarray}

It is interesting that the Higgs quartic coupling 
$\lambda\approx 20$ corresponding to the critical coupling $g_Q^c\approx 9.1$ in 
Fig.~\ref{figY} is very close to the above estimate. Adopting the relation 
$v=\sqrt{-\mu^2/\lambda}$ and the relevant numerical outcomes, we do observe 
the satisfactory consistency $\sqrt{2\lambda}=6.2\approx \sqrt{-\tilde\mu^2}g_Q^c=6.3$, which 
supports our analysis. The quartic coupling then drops to $\lambda\approx 0.1$ 
quickly with the decrease of the energy scale as displayed in Fig.~\ref{figY}. We point out 
that the fixed-point value of $\lambda$ remains as $\lambda\approx 17$, even when its value 
at the symmetry restoration scale is set to zero; the evolutions of the other couplings will 
generate a nonvanishing $\lambda$, which always approaches the fixed-point value 
asymptotically.

Once the heavy quark condensates are created, they induce the negative quadratic term and
the positive quartic term from Eq.~(\ref{lam}), which specify a nontrivial minimum of the 
Higgs potential at $\phi\approx v$, resulting in the dynamical EWSB. Fourth generation leptons 
and all particles in the SM then gain their masses through the VEV, and the SM is brought 
into the symmetry broken phase. The masses in the SM evolve to their values at the 
electroweak scale under the governance of the RG equations \cite{Hung:2009hy} after the 
EWSB. The critical Yukawa coupling $g_Q^c\approx 9.1$ corresponds to a quark mass 
1.6 TeV, which is above the critical fermion mass 1.23 TeV for forming bound states in an
attractive Yukawa potential \cite{Hung:2009hy}. Their effect on the EWPT will be inspected
in the next section. Fourth generation leptons and other SM fermions such as top quarks 
do not form bound states.

\section{The first-order phase transition}
\label{fir}

\subsection{Effective Higgs potential}
\label{31}

As stated in the Introduction, the SM contains the sufficiently strong source for the 
baryon number violation associated with the triangle anomaly \cite{tHooft:1976rip}. 
It originates from the vacuum structure of nonablian gauge theories, i.e., the gauge 
field configurations with minimal energies. These vacua, separated by energy barriers 
of height proportional to the VEV $v$, carry different integer Chern-Simons numbers or
baryon numbers. The tunneling between various vacua via sphaleron transitions then 
violates the baryon number (and also the lepton number). Although the tunneling is 
seriously damped at zero temperature, the transitions occurring at finite $T\gtrsim 100$ 
GeV make possible hopping over the barriers \cite{Kuzmin:1985mm}. As to another Sakharov's 
criterion related to the $CP$ violation, it has been widely recognized that the KM 
mechanism in the SM is not enough to account for the BAU. We will argue that the $CP$ 
violation can be notably enhanced by a sequential fourth generation in the SM4 at the end 
of this section. Here we concentrate on the strength of the EWPT in the SM4.

It has been known that the EWPT in the SM is not strongly first-order. To achieve the 
first-order EWPT and realize the electroweak baryogenesis, the finite-temperature effective 
Higgs potential must be substantially modified by introducing new degrees of freedom with 
large couplings to Higgs bosons. Whether the EWPT can be strongly first-order in the 
presence of extra heavy fermions has been surveyed extensively 
\cite{Ham:2004xh,Carena:2004ha,Fok:2008yg,Kikukawa:2009mu}, and negative conclusions 
were drawn. We will revisit the impact of fourth generation fermions in the 
specific mass pattern, predicted in our dispersive formalism, on the strength of the 
EWPT. As in \cite{Kikukawa:2009mu}, we construct the effective potential 
$V_{\rm eff}(\phi,T,\mu_R)$ up to one loop,
\begin{eqnarray}
V_{\rm eff}(\phi,T,\mu_R)=V_0(\phi)+V_1(\phi,\mu_R)+V_{T}(\phi,T),\label{efv}
\end{eqnarray}
where $V_0$ is the tree-level zero-temperature potential at the 
electroweak scale, $V_1$ is the one-loop Coleman-Weinberg potential \cite{Coleman:1973jx} 
with the renormalization scale $\mu_R$, and $V_{T}$ is the one-loop finite-temperature 
potential \cite{Dolan:1973qd}. The renormalization scale $\mu_R$ is below the symmetry 
restoration scale, so the discussion is not subject to the compositeness condition of the 
strong sector defined at a high energy \cite{Kikukawa:2009mu}. 

The tree-level potential 
\begin{eqnarray}
V_0(\phi)=\frac{\mu^2}{2}\phi^2+\frac{\lambda}{4}\phi^4, \label{v0}
\end{eqnarray}
designates the VEV $v=\sqrt{-\mu^2/\lambda}$ with the coefficients 
$\mu^2=-m_H^2/2$ and $\lambda=-\mu^2/v^2$ in terms of the Higgs boson mass 
$m_H=125$ GeV \cite{PDG}. The above expression is a consequence of the RG evolution of 
the potential characterized by Eqs.~(\ref{sdu1}) and (\ref{lam}) from the symmetry 
restoration scale down to the electroweak scale.

\subsection{Heavy scalar contribution}
\label{32}

The one-loop potential $V_1$ collects contributions from particles that couple to Higgs 
bosons. As mentioned before, superheavy fermions of masses above 1.23 TeV form bound 
states in a deep Yukawa potential \cite{Hung:2009hy}, so fourth generation quarks, with 
$m_{t'}\approx 200$ TeV and $m_{b'}= 2.7$ TeV, appear as heavy scalars (or pseudoscalars) 
at the electroweak scale. To address the coupling between Higgs bosons 
and the heavy scalar denoted by $\eta$, we propose the effective interaction 
$\lambda'\phi^2\eta^2/2$ motivated by typical terms in a two-Higgs doublet model 
\cite{Kikukawa:2009mu}. The strength $\lambda'$ will be determined in this subsection. 
Consider the Higgs boson scattering process $H(p)H(0)\to H(0)H(p)$, where the momenta of 
the colliding Higgs bosons are labeled in Fig.~\ref{fig2}(a). This kinematic arrangement 
simplifies the loop calculation as seen shortly. For a momentum $p$ much higher than the 
electroweak scale, the scattering is described by the box diagram in Fig.~\ref{fig2}(a) 
with internal heavy quarks $Q$. For a momentum $p$ around the electroweak scale, the 
scattering is described effectively by Fig.~\ref{fig2}(c) with a loop of the heavy scalar 
$\eta$. Figure~\ref{fig2}(c) can be also plotted in the $t$ channel, which contributes 
equally under the special assignment of the external momenta. 

We deduce the coupling $\lambda'$ by matching the contribution of 
Fig.~\ref{fig2}(c) to that of Fig.~\ref{fig2}(a) in the high momentum region. Both 
diagrams with massive internal particles exist in the symmetry broken phase, so
the vertex in Fig.~\ref{fig2}(c) corresponds to $\lambda'h^2\eta^2/2$ actually with the
physical Higgs field $h$, which is one of the pieces in $\lambda'\phi^2\eta^2/2$ under 
the transformation $\phi=h+v$. To obtain $\lambda'$, it suffices to focus on this piece.
A remark on the role of Fig.~\ref{fig2}(b) is in order. Figures~\ref{fig2}(a) and 
\ref{fig2}(b) stand for a perturbative contribution and a power correction, respectively, 
in the operator product expansion of the four-point correlation function defined by scalar 
currents. With the large external momentum $p$, the denominator $m^2(0)$ in Eq.~(\ref{la0}) 
would be replaced by $p^2$, making explicit the power suppression. Hence, 
Fig.~\ref{fig2}(b) is not included in the matching procedure. After deriving 
$\lambda'$, we use it to evaluate the heavy scalar contributions to $V_1$ and $V_T$.

The loop integral for Fig.~\ref{fig2}(a) with internal heavy quarks $Q$ is written as
\begin{eqnarray}
\Pi_Q(s)=-N_c\left(\frac{g_Q}{\sqrt{2}}\right)^4i\int \frac{d^4l}{(2\pi)^4}
\frac{{\rm tr}[(\not l-\not p+m_Q)(\not l-\not p+m_Q)(\not l+m_Q)(\not l+m_Q)]}
{[(l-p)^2-m_Q^2]^2(l^2-m_Q^2)^2}\label{int}
\end{eqnarray}
with $s=p^2$ and the minus sign coming with the quark loop. 
The simpler integrand for the box diagram is attributed to the special choice of 
the external momenta. A straightforward computation of Eq.~(\ref{int}) turns in
\begin{eqnarray}
\Pi_Q(s)&=&-\frac{N_c g_Q^4}{16\pi^2}\int_0^1du\left\{\ln\left[1-u(1-u)\frac{s}{m_Q^2}\right]
+\cdots\right\},\label{4a}
\end{eqnarray}
where $u$ is the Feynman parameter, the ultraviolet pole 
has been regularized, and the renormalization scale is set to $m_Q$.
We retain only the leading logarithmic term at large $s$, and keep in mind that
our result suffers the theoretical uncertainty from the regularization scheme.

The effective diagram in Fig.~\ref{fig2}(c) with a heavy scalar loop 
gives 
\begin{eqnarray}
\Pi_\eta(s)&=&[2\lambda'(s)]^2i\int \frac{d^4l}{(2\pi)^4}
\frac{1}{[(l-p)^2-m_\eta^2](l^2-m_\eta^2)},
\end{eqnarray}
where the scale-dependent coupling $2\lambda'(s)$ is from the $h^2\eta^2$ 
vertex, and $m_\eta$ is the heavy scalar mass. It is trivial to isolate the leading 
logarithmic term from the above integral,
\begin{eqnarray}
\Pi_\eta(s)
=-\frac{\lambda'(s)^{2}}{4\pi^2}
\int_0^1du\ln\frac{m_Q^2}{m_\eta^2-u(1-u)s},\label{4c}
\end{eqnarray}
with the renormalization scale being set to $m_Q$ as well.

Equating Eq.~(\ref{4c}) with Eq.~(\ref{4a}), i.e., 
$\Pi_\eta(s)=\Pi_Q(s)$, at high $s$, we fix $\lambda'(s)$, and then acquire the effective 
coupling at the electroweak scale $s=v^2$,
\begin{eqnarray}
\lambda'(v^2)\approx \frac{g_Q^2v}{2m_Q\sqrt{\ln(m_\eta^2/m_Q^2)}}
=\frac{g_Q}{2\sqrt{\ln[2m_\eta^2/(g_Q^2v^2)]}}.\label{eff}
\end{eqnarray}
The expansion $\ln(1-x)\approx -x$ for small $x$ has been applied, and the relation 
$m_Q=\sqrt{2}g_Q/v$ has been inserted to arrive at the final expression.
As shown in \cite{Li:2023fim}, fourth generation quarks may form more than one bound 
state under the Yukawa potential. Viewing the suppression of the effective 
coupling by the heavy bound state mass in Eq.~(\ref{eff}),  
we consider only the lowest lying state for Fig.~\ref{fig2}(c). 
Precisely speaking, the ground state of fourth generation quarks is a pseudoscalar, 
instead of a scalar. It has been found through relativistic analyses 
\cite{Ikhdair:2012zz} that the lightest pseudoscalar formed by the $b'\bar b'$ quark 
pair with $m_{b'}= 2.7$ TeV has a mass $3.23$ TeV \cite{Li:2023fim}. We thus have, for 
the inputs of the critical coupling $g_Q^c= 9.1$ and $m_\eta\approx 3.23$ TeV,
\begin{eqnarray}
\lambda'\approx 3.8.\label{lam1}
\end{eqnarray}

\begin{figure}[tbp]
\centering 
\includegraphics[width=.35\textwidth]{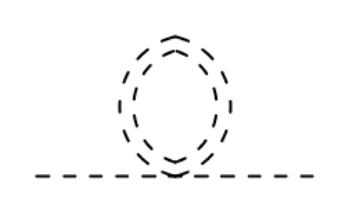}

\caption{\label{fey}
Contribution to the one-loop Coleman-Weinberg potential $V_1$
from a heavy scalar $\eta$ (denoted by a double dashed line) at the power of $\phi^2$.}
\end{figure}

The contributions from heavy scalars $\eta$ to the one-loop potential $V_1$ 
arise from the diagrams composed of the effective interaction $\phi^2\eta^2$, i.e.,
Fig.~\ref{fey} proportional to $\phi^2$ and Fig.~\ref{fig2}(c) proportional to $\phi^4$, 
where the external legs represent the field $\phi$. Those yielding higher powers of 
$\phi^2$ can be drawn in a similar way. As demonstrated later, the $O(1)$ effective 
coupling in Eq.~(\ref{lam1}) at the electroweak scale, amplifying the scalar loop 
corrections to $V_1$, is crucial for generating the first-order EWPT. We will 
corroborate that our observation for the ratio $\phi_c/T_c$ is insensitive to the 
variation of $\lambda'$ around the estimate $\lambda'\approx 3.8$. It is 
also worthwhile to study the effective Lagrangian which describes the complete 
Higgs-heavy-scalar interactions and its phenomenological impacts.

\subsection{Strength of electroweak phase transition}
\label{33}

The $SU_C(3)\times SU_L(2)\times U_Y(1)$ gauge interactions are negligible in the current
framework \cite{Kikukawa:2007zk}, since the gauge couplings are smaller than the effective 
coupling $\lambda'$ in Eq.~(\ref{lam1}), and than the Yukawa couplings $g_t$, $g_L$ 
and $g_{\nu_4}$ associated with a top quark, a fourth generation charged lepton $L$ and a 
fourth generation neutrino $\nu_4$, respectively. 
The one-loop Coleman-Weinberg potential reads \cite{Coleman:1973jx}
\begin{eqnarray}
V_1(\phi,\mu_R)=\frac{1}{64\pi^2}\sum_{i=H,\eta,t,L,\nu_4} 
n_i m_i^4(\phi)\left[\ln\frac{m_i^2(\phi)}{\mu_R^2}
-\frac{3}{2}\right]+\frac{1}{2}A(\mu_R)\phi^2,\label{v1p}
\end{eqnarray}
where $n_H=1$, $n_\eta=1$, $n_{t}=-12$, $n_L=-4$ and $n_{\nu_4}=-4$ are the degeneracies 
per particle, and $m_H^2(\phi)=\mu^2+a_H\phi^2$, $m_\eta^2(\phi)=a_\eta\phi^2$,
$m_{t}^2(\phi)=a_{t}\phi^2$, $m_{L}^2(\phi)=a_{L}\phi^2$
and $m_{\nu_4}^2(\phi)=a_{\nu_4}\phi^2$ with $a_H=3\lambda$, 
$a_\eta=\lambda'$, $a_{t}=g_{t}^2/2$, $a_{L}=g_{L}^2/2$ and $a_{\nu_4}=g_{\nu_4}^2/2$, 
respectively, are the field-dependent masses. 
The one-loop potential shifts the minimum away from the tree-level location $\phi=v$, 
which can be preserved by imposing the renormalization condition on the quadratic term of the
effective potential. This is accomplished by the second term in Eq.~(\ref{v1p})
\cite{Kikukawa:2009mu},
\begin{eqnarray}
A(\mu_R)=-\frac{1}{16\pi^2}\sum_{i=H,\eta,t,L,\nu_4} n_i a_i m_i^2(v)
\left[\ln\frac{m_i^2(v)}{\mu_R^2}-1\right],
\end{eqnarray}
that enforces the minimum of $V_0(\phi)+V_1(\phi,\mu_R)$ to be at $\phi=v$.

The one-loop finite-temperature effective potential is written as 
\begin{eqnarray}
V_{T}(\phi,T)=\frac{T^4}{2\pi^2}\left[\sum_{i=H,\eta} n_iJ_B(m_i^2(\phi)/T^2)+
\sum_{i=t,L,\nu_4}n_{i}J_F(m_i^2(\phi)/T^2)\right],\label{vt}
\end{eqnarray}
with the thermal functions
\begin{eqnarray}
J_{B,F}(x)=\int_0^\infty dyy^2\ln\left[1\mp\exp\left(-\sqrt{y^2+x}\right)\right].
\label{ther}
\end{eqnarray}
The contributions of the $T$-dependent ring diagrams \cite{Dolan:1973qd,Weinberg:1974hy,
Anderson:1991zb}, improving the reliability of the perturbative expansion near the critical 
temperature, are incorporated by adding
\begin{eqnarray}
\Pi_H(T)=\left(3\lambda+\lambda'+3g_t^2+g_L^2+g_{\nu_4}^2\right)\frac{T^2}{12},\;\;\;\;
\Pi_\eta(T)=\lambda'\frac{T^2}{12},
\end{eqnarray}
to $m_H^2(\phi)$ and $m_\eta^2(\phi)$, respectively, in $V_1(\phi,\mu_R)$ and in 
$V_{T}(\phi,T)$ \cite{Kikukawa:2009mu,Curtin:2016urg}. Their effect, smoothing the low-$\phi$ 
behavior of $V_{\rm eff}(\phi,T,\mu_R)$, helps the identification of the 
trivial minimum at $\phi=0$.

\begin{figure}[tbp]
\centering 
\includegraphics[width=.45\textwidth]{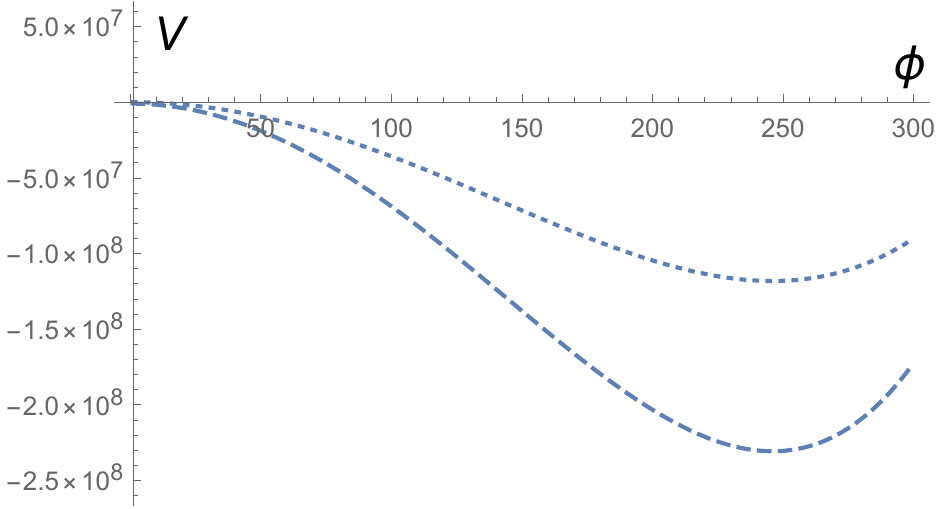}
\hfill
\includegraphics[width=.45\textwidth]{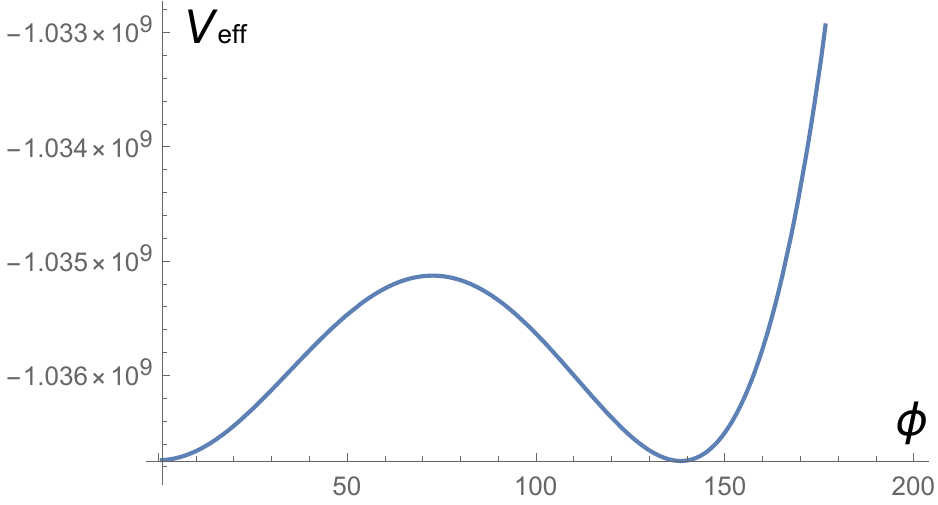}

(a)\hspace{8.0cm}(b)

\caption{\label{fig5}
(a) Dependencies of $V_0$ (dotted line) and $V_0+V_1$ 
(dashed line) in units of GeV$^4$ on $\phi$ in units of GeV. 
(b) Dependence of $V_{\rm eff}$ in units of GeV$^4$ on $\phi$ in 
units of GeV for $T_c=148.8$ GeV.}
\end{figure}

The renormalization scale is set to the electroweak scale $\mu_R=v$, which 
is the characteristic scale for the considered effective Higgs potential. 
The Yukawa couplings take the values $g_{t,L,\nu_4}=\sqrt{2}m_{t,L,4}/v$ with the top 
quark mass $m_t=173$ GeV \cite{PDG}. The dependencies of $V_0(\phi)$ and 
$V_0(\phi)+V_1(\phi,v)$, and $V_{\rm eff}(\phi,T_c,v)$ in Eq.~(\ref{efv}) on the field 
$\phi$ are depicted in Figs.~\ref{fig5}(a) and \ref{fig5}(b), respectively, for the critical 
temperature $T_c=148.8$ GeV. It is noticed that the heavy scalars from fourth generation 
quarks contribute to $V_1$ dominantly owing to the sizable coupling $\lambda'$, which 
deepens the valley of $V_0$ at $\phi=v$ as indicated in the plot. This modification builds 
a barrier between the trivial vacuum at $\phi=0$ and the nontrivial vacuum at $\phi=\phi_c$ 
in Fig.~\ref{fig5}(b), when the latter is lifted up to the same height as of the former by 
increasing the temperature. We point out that the heavy scalars  
magnify the cubic term $m_i^3(\phi)/T^3$ \cite{Dolan:1973qd,Patel:2011th} in the power 
expansion of the thermal function $J_B(m_i^2(\phi)/T^2)$ in Eq.~(\ref{vt}). The slope 
of $V_T$ in $\phi$ grows with the temperature. The contribution from fourth generation 
leptons enhances this slope, such that a lower $T_c$ is needed for the lift-up compared to 
the case without fourth generation leptons. Figure~\ref{fig5}(b) for the effective 
potential $V_{\rm eff}(\phi,T_c,v)$ reveals two degenerate minima 
$V_{\rm eff}(0,T_c,v) = V_{\rm eff}(\phi_c, T_c,v)$ residing at $\phi= 0$ and 
$\phi=\phi_c=138.4$ GeV, leading to the ratio $\phi_c/T_c\approx 0.9$.  
The EWPT in the SM4 is then likely first-order, for the criterion 
$\phi_c/T_c\gtrsim 1$ is roughly met. The critical temperature $T_c=148.8$ GeV is also
consistent with the expectation $T_c\gtrsim 100$ GeV for the EWPT stated in Sec.~\ref{31}.

We examine the sensitivity of the numerical outcomes to the variation of the effective 
quartic coupling $\lambda'$, prompted by the regularization scheme 
dependence mentioned in the previous subsection; the increase (decrease) of $\lambda'$ 
by 20\% reduces $\phi_c/T_c$ by about 3\% (4\%) with $\phi_c=145.4$ GeV and $T_c=161.1$ GeV
($\phi_c=124.6$ GeV and $T_c=139.9$ GeV). 
It happens that Eq.~(\ref{lam1}) gives a ratio $\phi_c/T_c$ very close to its maximum in 
our theoretical setup, explaining the stability of the results. Turning off the contribution 
from fourth generation leptons produces $\phi_c=148.2$ GeV and $T_c=177.2$ GeV, 
i.e., $\phi_c/T_c\approx 0.8$. That is, the strength of the EWPT is weakened a bit,
evincing the small influence from fourth generation leptons. Deleting the heavy scalar 
contribution diminishes the barrier between the two vacua, and renders the EWPT  
second-order or crossover. It highlights that superheavy fourth generation quarks 
offer a necessary ingredient to the effective potential through their bound states. 
The postulation that fourth generation fermions cannot strengthen the EWPT 
\cite{Fok:2008yg} is then bypassed. 

The field-dependent mass $m_\eta^2(\phi)=a_\eta\phi^2$ in Eq.~(\ref{v1p}) 
implies that the $\eta$ mass is also induced by the electroweak symmetry 
breaking, since a heavy scalar is formed after the symmetry breaking. This fact should 
be respected in the construction of the effective Lagrangian for Higgs-heavy-scalar 
interactions, in which the quadratic term $\eta^2$ is absent at tree level with the 
coefficient $\mu^2_\eta=0$. One can then derive the effective potential $V_{\rm eff}(\eta)$ 
for the heavy scalar $\eta$ in the same manner, assuming the existence of a quartic term 
$\eta^4$, whose strength is not crucial for our reasoning. The effective couplings between 
heavy scalars and other light fermions are expected to be small, so the fermion contributions 
are neglected for simplicity. Owing to $\mu^2_\eta=0$, there exists only the trivial vacuum 
at $\eta=0$ in the tree-level potential $V_0(\eta)$, which is preserved under the one-loop
Coleman-Weinberg correction $V_1(\eta)$. It can be verified easily that 
this feature remains when the one-loop finite-temperature potential $V_T(\eta)$ is included; 
the shape of $V_{\rm eff}(\eta)$ is not altered much under the variations of the 
renormalization scale $\mu_R$ and the temperature $T$. That is, the heavy scalar $\eta$ does 
not acquire a non-zero VEV throughout the thermal history, and the results obtained in the 
present work will not be impacted by the heavy scalar potential.

\subsection{Enhancement of $CP$ violation}

As commented before, the KM mechanism in the SM is not enough to account for the BAU.
The conventional argument proceeds with the Jarlskog invariant 
\cite{Jarlskog:1985ht,Jarlskog:1985cw}, which involves a significant suppression by the 
ratio of a quark mass over the critical temperature to the twelfth power 
\cite{Shaposhnikov:1986jp}; the magnitude of the relevant $CP$ violation 
in the SM depends on quark masses via 
\begin{eqnarray}
J=(m_t^2-m_c^2)(m_t^2-m_u^2)(m_c^2-m_u^2)(m_b^2-m_s^2)(m_b^2-m_d^2)(m_s^2-m_d^2)A_{123},
\label{J1}
\end{eqnarray}
where $A_{123}$ is twice the area of any triangle formed by the unitarity condition of the
CKM matrix $V_{\rm CKM}^\dagger V_{\rm CKM} = I$. Equation~(\ref{J1}) leads to a 
dimensionless quantity
\begin{eqnarray}
\frac{J}{T_c^{12}}\sim \frac{J}{(150\; {\rm GeV})^{12}} \sim 10^{-22},\label{dim}
\end{eqnarray}
which is consistent with the observation in \cite{Gavela:1994dt}. The critical temperature
$T_c\approx 150$ GeV is quoted from the derivation in Sec.~\ref{33}.
It is obvious that this value is too small to accommodate the asymmetry between matter and 
antimatter in terms of the baryon-to-photon ratio
\begin{eqnarray}
\eta_B \equiv\frac{n_B-n_{\bar B}}{n_\gamma}\sim 10^{-10}.\label{eta}
\end{eqnarray}
It thus demands additional prominent sources of $CP$ violation for the baryogenesis.

We will elaborate qualitatively that the $CP$ violation can be intensified remarkably by 
including sequential fourth generation quarks \cite{Hou:2008xd,Hou:2010wf}. The quantitative 
study on the amount of $CP$ violation for producing the observed BAU is intricate, to which 
on-going efforts are being devoted \cite{Kharzeev:2019rsy,Chao:2019smr,Basler:2021kgq,
Jangid:2023jya,Kainulainen:2024qpm,Liu:2024mdo,Giovanakis:2024rvg}.
Take the $sb$ element of $V_{\rm CKM} V_{\rm CKM}^\dagger$ associated with the $b\to s$ 
transition as an example, 
\begin{eqnarray}
V_{us}V^*_{ub} + V_{cs}V^*_{cb} + V_{ts}V^*_{tb} = 0,\label{tr1}
\end{eqnarray}
which is one of the triangles appearing in Eq.~(\ref{J1}).
The tininess of Eq.~(\ref{dim}) originates mainly from the small masses, 
$m_s^2 m_c^2 m_b^4/T_c^8$, rather than from the area $A_{123}$. A triangle in the SM 
changes to a quadrangle in the SM4. An analog to $J$ in the SM4 is written as 
\cite{Hou:2010wf} 
\begin{eqnarray}
J_{234}=(m_{t'}^2-m_t^2)(m_{t'}^2-m_c^2)(m_t^2-m_c^2)
(m_{b'}^2-m_b^2)(m_{b'}^2-m_s^2)(m_b^2-m_s^2)A_{234},
\label{J2}
\end{eqnarray}
where $A_{234}$ denotes the area of the quadrangle
\begin{eqnarray}
V_{us}V^*_{ub} + V_{cs}V^*_{cb} + V_{ts}V^*_{tb}+ V_{t's}V^*_{t'b}= 0.\label{tr2}
\end{eqnarray}
Because of the smallness of the magnitudes $|V_{us}V^*_{ub}|$ and 
$|V_{t's}V^*_{t'b}|$ \cite{Li:2024xnl}, the area $A_{234}$ is of the same order 
of $A_{123}$. 

We have verified that the heavy quark condensates are formed at the critical coupling 
$g_Q^c\approx 9.1$. The quartic coupling of the tree-level potential $V_0$ in Eq.~(\ref{v0})
then approaches its SM value $\lambda\approx 0.1$ immediately 
under the RG running as seen in Fig.~\ref{figY}. The other 
fermion Yukawa couplings, represented by the top quark one in Fig.~\ref{figY}, have 
also reached their SM values approximately, when the heavy quark condensates are 
established. Hence, we are allowed to assess the strength of the $CP$ violation
for the baryogenesis using the quark masses $m_{t'}=m_{b'}=1.6$ TeV (corresponding to 
$g_Q^c= 9.1$), $m_t=173$ GeV, $m_b= 4$ GeV, $m_c=1.5$ GeV and $m_s= 0.1$ GeV. It is 
trivial to attain that the Jarlskog invariant in 
the SM4 is larger than the one in the SM by a factor
\begin{eqnarray}
\frac{J_{234}}{J}\approx \left(\frac{m_{t'}^2m_{b'}^2}{m_tm_cm_bm_s}\right)^2\approx 10^{21},
\end{eqnarray}
which marks a magnificent enhancement of the $CP$ violation. The ratio $\eta_B$ is 
about $10^{-3}$ times $J_{234}/T_c^{12}$ \cite{Gavela:1994dt}, i.e., $\eta_B\sim 10^{-4}$.
It hints that the BAU might be explained in the SM4, though a more quantitative and
rigorous investigation needs to be conducted before a solid conclusion can be drawn.

\section{Conclusion}
\label{con}

The SM4 with sequential fourth generation fermions, as a natural and simple extension 
of the SM, deserves thorough exploration. This work discusses the dynamical aspects 
on the electroweak symmetry breaking and phase transition realized in the SM4. We have 
certified the existence of a symmetric phase near the ultraviolet fixed point of the two-loop 
RG evolution in the SM4; the critical Yukawa coupling $g_Y^c\approx 8.0$ for fourth generation 
fermions to form condensates is greater than their common fixed-point value $g_Y^f\approx 7$.
The restoration of the electroweak symmetry at a high energy consolidates  
the dispersive constraints on the fermion masses and mixing angles in our previous works. 
As the energy scale decreases, the Yukawa coupling of fourth generation quarks (leptons) 
ascends (descends) from the fixed-point value. When the former exceeds the critical coupling
$g_Q^c\approx 9.1$, the heavy quark condensation is triggered, which breaks the electroweak 
symmetry, and particles in the SM4 gain masses. The above scenario has been confirmed 
by solving the coupled DS equations for the fermion and Higgs boson masses. It has been also 
shown that the heavy quark condensates induce not only the quadratic, but also quartic 
terms in the Higgs potential, and that their coefficients gives rise to the VEV $v$.

Employing the standard formulation for the EWPT, we analyzed the effects from
fourth generation fermions on the Coleman-Weinberg potential $V_1$ and the one-loop 
finite-temperature potential $V_T$. Fourth generation quarks, with their masses
above a TeV scale, form bound states under the huge Yukawa interaction. The 
coupling between the bound states and Higgs bosons was determined by 
matching the effective diagram for the Higgs scattering process
with a heavy scalar loop in Fig.~\ref{fig2}(c) to the box diagrams with internal 
fourth generation quarks in Fig.~\ref{fig2}(a). The loop corrections to 
the effective Higgs potential resulting from heavy scalars have been displayed up
to the power $\phi^4$ in Fig.~\ref{fig2}(c) and Fig.~\ref{fey}. Their contributions 
to $V_1$ take the same form as those from Higgs bosons but with a different coupling. 
Their impact deepens the nontrivial vacuum of the effective Higgs potential around $\phi=v$,
such that a barrier between the trivial vacuum $\phi=0$ and the nontrivial one is erected
as the temperature grows. Fourth generation leptons are not responsible for the formation 
of the condensates. They increase the slope of the finite-temperature potential, so that 
the two vacua $\phi=0$ and $\phi=\phi_c=138.4$ GeV become degenerate at 
the lower critical temperature $T_c=148.8$ GeV. We read off the ratio 
$\phi_c/T_c\approx 0.9$, which should suffice for producing the first-order EWPT. Along 
with the strong sphaleron transitions inherent in the SM and the magnified $CP$ violation 
source from fourth generation quarks, we conclude that the electroweak baryogenesis can 
be achieved.



\acknowledgments

We thank C.W. Chiang, Y. Chung, W.S. Hou, V.Q. Tran, P.Y. Tseng, T.C. Yuan, X.B. Yuan and 
M.R. Wu for fruitful discussions. This work was supported in part by National Science and 
Technology Council of the Republic of China under Grant No. NSTC-113-2112-M-001-024-MY3.



\end{document}